 \documentclass[12pt]{article}

 \usepackage{hyperref}
 \usepackage{color}

 \usepackage{graphicx}
 \usepackage{cite}

 \usepackage{longtable} 

 \long\def\begincomment#1\endcomment{}

\begin{document}


\begin{center}
{\LARGE\bf Leptonic decay-constant ratio $f_K/f_\pi$ from\\
lattice QCD using 2+1 clover-improved\\[2mm]
fermion flavors with 2-HEX smearing}
\end{center}

\vspace{10pt}

\begin{center}
Stephan~D\"urr$\,^{a,b}$ \footnote{\tt durr (AT) itp.unibe.ch},
Zolt\'an~Fodor$\,^{a,b,c}$,
Christian~Hoelbling$\,^{a}$,
Stefan~Krieg$\,^{a,b}$,
Laurent~Lellouch$\,^{d}$,
Thomas~Lippert$\,^{a,b}$,
Thomas~Rae$\,^{a}$,
Andreas~Sch\"afer$\,^{e}$,
Enno~E.~Scholz$\,^{e}$ \footnote{\tt enno.scholz (AT) physik.uni-regensburg.de},
K\'alm\'an~K.~Szab\'o$\,^{a,b}$,
Lukas~Varnhorst$\,^{a}$
\\[10pt]
${}^a${\sl University of Wuppertal, Gau{\ss}stra{\ss}e\,20, D-42119 Wuppertal, Germany}\\
${}^b${\sl J\"ulich Supercomputing Centre, Forschungszentrum J\"ulich, D-52425 J\"ulich, Germany}\\
${}^c${\sl Institute for Theoretical Physics, E\"otv\"os University, H-1117 Budapest, Hungary}\\
${}^d${\sl CNRS, Aix Marseille U., U.\ de Toulon, CPT, UMR~7332, F-13288 Marseille, France}\\
${}^e${\sl University of Regensburg, Universit\"atsstra{\ss}e 31, D-93053 Regensburg, Germany}
\end{center}

\vspace{10pt}

\begin{abstract}
  \noindent
   We present a calculation of the leptonic decay-constant ratio
   $f_K/f_\pi$ in 2+1 flavor QCD. Our data set includes five lattice
   spacings and pion masses reaching down below the physical
   one. Special emphasis is placed on a careful study of all
   systematic uncertainties, especially the continuum extrapolation.
   Our result is perfectly compatible with the first-row unitarity
   constraint of the Standard Model.
\end{abstract}

\vspace{10pt}


\newcommand{\pad}{\partial}
\newcommand{\hqu}{\hbar}
\newcommand{\til}{\tilde}
\newcommand{\pri}{^\prime}
\renewcommand{\dag}{^\dagger}
\newcommand{\<}{\langle}
\renewcommand{\>}{\rangle}
\newcommand{\gaf}{\gamma_5}
\newcommand{\nab}{\nabla}
\newcommand{\lap}{\triangle}
\newcommand{\dal}{{\sqcap\!\!\!\!\sqcup}}
\newcommand{\trc}{\mathrm{tr}}
\newcommand{\Trc}{\mathrm{Tr}}
\newcommand{\Mpi}{M_\pi}
\newcommand{\Fpi}{F_\pi}
\newcommand{\Mka}{M_K}
\newcommand{\Fka}{F_K}
\newcommand{\Met}{M_\et}
\newcommand{\Fet}{F_\et}
\newcommand{\Mss}{M_{\bar{s}s}}
\newcommand{\Fss}{F_{\bar{s}s}}
\newcommand{\Mcc}{M_{\bar{c}c}}
\newcommand{\Fcc}{F_{\bar{c}c}}
\newcommand{\fpi}{f_\pi}
\newcommand{\fka}{f_K}

\newcommand{\al}{\alpha}
\newcommand{\be}{\beta}
\newcommand{\ga}{\gamma}
\newcommand{\de}{\delta}
\newcommand{\ep}{\epsilon}
\newcommand{\ve}{\varepsilon}
\newcommand{\ze}{\zeta}
\newcommand{\et}{\eta}
\renewcommand{\th}{\theta}
\newcommand{\vt}{\vartheta}
\newcommand{\io}{\iota}
\newcommand{\ka}{\kappa}
\newcommand{\la}{\lambda}
\newcommand{\rh}{\rho}
\newcommand{\vr}{\varrho}
\newcommand{\si}{\sigma}
\newcommand{\ta}{\tau}
\newcommand{\ph}{\phi}
\newcommand{\vp}{\varphi}
\newcommand{\ch}{\chi}
\newcommand{\ps}{\psi}
\newcommand{\om}{\omega}

\newcommand{\bdm}{\begin{displaymath}}
\newcommand{\edm}{\end{displaymath}}
\newcommand{\bea}{\begin{eqnarray}}
\newcommand{\eea}{\end{eqnarray}}
\newcommand{\beq}{\begin{equation}}
\newcommand{\eeq}{\end{equation}}

\newcommand{\mr}{\mathrm}
\newcommand{\mb}{\mathbf}
\newcommand{\ri}{\mr{i}}
\newcommand{\Nf}{N_{\!f}}
\newcommand{\Nc}{N_{ c }}
\newcommand{\Nt}{N_{ t }}
\newcommand{\MeV}{\,\mr{MeV}}
\newcommand{\GeV}{\,\mr{GeV}}
\newcommand{\fm}{\,\mr{fm}}
\newcommand{\MSbar}{\overline{\mr{MS}}}

\hyphenation{topo-lo-gi-cal simu-la-tion theo-re-ti-cal mini-mum con-tinu-um}


\section{Introduction\label{sec:intro}}


Leptonic decays of pseudoscalar mesons provide a convenient way to determine Cabibbo-Kobayashi-Maskawa (CKM) matrix elements in the Standard Model (SM) and may, in the future, give access to some Beyond Standard Model (BSM) processes.
Today comparison with CKM matrix elements obtained from transition form factors allows for testing the internal consistency of the SM.
Obviously, in this process the highest possible precision in both theory and experiment is crucial for eventually being able to see indications for New Physics.

In this paper, we provide a computation of the decay-constant ratio $\fka/\fpi$ in the isospin symmetric limit of QCD, i.e.\ with two degenerate light quarks which have the same mass as the average $\frac{1}{2}(m_\mr{u}^\mr{phys}+m_\mr{d}^\mr{phys})$ in nature.
By combining $\fka/\fpi$ with a factor taken from Chiral Perturbation Theory (ChPT) we determine the charged decay-constant ratio $f_{K^\pm}/f_{\pi^\pm}$.
The latter object connects to the ratio of the experimentally measured widths via the relation \cite{Marciano:2004uf}
\beq
\frac{\Gamma(K^\pm\to\ell\nu_\ell)}{\Gamma(\pi^\pm\to\ell\nu_\ell)}=
\frac{V_\mathrm{us}^2}{V_\mathrm{ud}^2}
\frac{f_{K^\pm}^2}{f_{\pi^\pm}^2}
\frac{M_{K^\pm}}{M_{\pi^\pm}}
\frac{(1-m_\ell^2/M_{K^\pm}^2)^2}{(1-m_\ell^2/M_{\pi^\pm}^2)^2}
\;(1\!+\!\de_\mr{em})
\label{eq:marciano}
\eeq
where $\ell=e^\pm,\mu^\pm$ and $\nu_\ell$ denotes the corresponding (electron or muon) neutrino or anti-neutrino.
Here and in the following we use the standard parameterization where the complex phase in the first row of the CKM matrix is assigned exclusively to $V_\mr{ub}$.
Marciano advocates this form, because some of the experimental uncertainties in the determination of $\Gamma(K\to\ell\nu_\ell), \Gamma(\pi\to\ell\nu_\ell)$, some of the lattice uncertainties in the computation of $\fka,\fpi$, and some of the radiative corrections in $\de_\mr{em}=\de_K-\de_\pi$ cancel.

Our lattice computation uses Wilson fermions \cite{Wilson:1974sk,Wilson:1975id} with a tree-level clover term \cite{Sheikholeslami:1985ij} and two levels of HEX smearing \cite{Capitani:2006ni} along with a Symanzik improved gauge action \cite{Luscher:1985zq}.
In total we use 47 ensembles with five different lattice spacings which cover a wide range of pion masses (approximately between $130\MeV$ and $680\MeV$) and kaon masses (such that in most cases the strange-quark mass is close to its physical value) in large boxes (such that finite-volume effects are subdominant).
As we use Wilson-type fermions, a welcome feature of the Marciano setup is that the factors $Z_A$ that would be needed to convert the bare decay constants into the physical $\fpi$ and $\fka$, respectively, cancel in the ratio given in Eq.~(\ref{eq:marciano}).

These ensembles have previously been used to determine the light quark masses \cite{Durr:2010vn,Durr:2010aw}, indirect CP violation \cite{Durr:2011ap}, some low-energy constants of QCD that appear in ChPT \cite{Durr:2013goa}, as well as the light and strange nucleon sigma terms \cite{Durr:2015dna}.
In the present work $\fka/\fpi$ is measured on each ensemble, and the goal is to use all these data for a controlled interpolation to the physical mass point, and extrapolation to zero lattice spacing ($a\to0$) and infinite volume ($L\to\infty$).
The philosophy of the analysis is that the parameterization of the data is achieved in a modular fashion, such that we have a factor for the dependence on the quark mass, one for the dependence on the lattice spacing $a$, and another one for the box size $L$.
For each factor several reasonable ans\"atze are considered, and the same statement holds true with respect to the cuts on the data that will be invoked (see below for details).
Overall we end up with $\mathcal{O}(1000)$ reasonable analyses (i.e.\ combined interpolations to the physical mass point and extrapolations $a\to0$ and $L\to\infty$).
Since each one of these is performed in a fully bootstrapped fashion, we can quote a reliable estimate of both the statistical and the systematic uncertainties.

The remainder of this paper is organized as follows.
In Sec.\,\ref{sec:ensembles} we give an overview of the ensembles used in this study.
We continue in Sec.\,\ref{sec:fitforms} with a discussion of the functional ans\"atze which are used to establish a combined interpolation to the physical mass point and extrapolation to zero lattice spacing and to infinite volume.
How these $\mathcal{O}(1000)$ analyses are distilled into a single number for $\fka/\fpi$ in the isospin limit of QCD is described in Sec.\,\ref{sec:results}.
This number is multiplied by a correction factor from ChPT to yield $f_{K^\pm}/f_{\pi^\pm}$, and upon combining the latter object with the PDG value $\Gamma(K\to\ell\nu_\ell)/\Gamma(\pi\to\ell\nu_\ell)$ and $\de_\mr{em}$ one finds $V_\mathrm{us}/V_\mathrm{ud}$ as described in Sec.\,\ref{sec:discussion}.
We conclude with a check of the CKM first-row unitarity property, based on the Hardy-Towner value of $V_\mathrm{ud}$, and compare our result to the literature.


\section{Overview of ensembles used\label{sec:ensembles}}


For our analysis, we use the 2-HEX ensembles generated by the Budapest-Marseille-Wuppertal Collaboration (BMW-Collab.) with $N_f=2+1$ flavors of tree-level clover-improved Wilson fermions with two HEX smearings and the tree-level Symanzik-improved gauge action \cite{Durr:2010vn,Durr:2010aw}.

\begin{table}[!tb]
\centering
\begin{tabular}{cccc}
\hline\hline
$\beta$ & $a^{-1}/\mr{GeV}$ & $\alpha_{N_f=3}$ & $\alpha_{N_f=4}$ \\ 
\hline\hline
3.31 & 1.670(07) & 0.327 & 0.333 \\ 
3.50 & 2.134(15) & 0.286 & 0.295 \\ 
3.61 & 2.576(28) & 0.262 & 0.271 \\ 
3.70 & 3.031(32) & 0.244 & 0.254 \\ 
3.80 & 3.657(37) & 0.227 & 0.237 \\ 
\hline\hline
\end{tabular}
\caption{\label{tab:aInv}
Lattice scale $a^{-1}$ and strong-coupling parameter $\alpha$ for each gauge coupling $\beta$. The strong-coupling parameter $\alpha$ is given in the $\MSbar$ scheme at scale $a^{-1}$ using both $\Nf=3$ and $\Nf=4$ matching; in our analysis we always use the average value of these two methods.}
\end{table}

The two light (up- and down-) quark flavors are mass-degenerate, with their common mass $m_\mr{ud}$ chosen to result in pion masses between approx.\ 130 and 680 MeV.
The single strange-quark mass $m_\mr{s}$ is taken close to the respective physical value in 45 ensembles and significantly heavier than $m_\mr{s}^\mr{phys}$ in two ensembles.
The ensembles are generated at five different gauge couplings ($\beta=3.31$, 3.50, 3.61, 3.70, and 3.80), and this results in lattice scales $a^{-1}$ between about $1.7\GeV$ and $3.7\GeV$.
These mass-independent lattice scales (i.e.\ determined for each set of ensembles at a fixed gauge coupling from the mass of the $\Omega$-baryon at the physical mass point as described in Refs.~\cite{Durr:2010vn,Durr:2010aw}) are collected in Tab.~\ref{tab:aInv}.
In total, we have 47 different ensembles at our disposal, with particulars given in Tab.~\ref{tab:ensDetails} of the Appendix.
For more information on the ensembles and how the meson masses and decay constants are extracted from the usual two-point correlators see Refs.~\cite{Durr:2010vn,Durr:2010aw,Durr:2013goa}.

\begin{figure}[!tb]
\centering
\includegraphics[width=.65\textwidth]{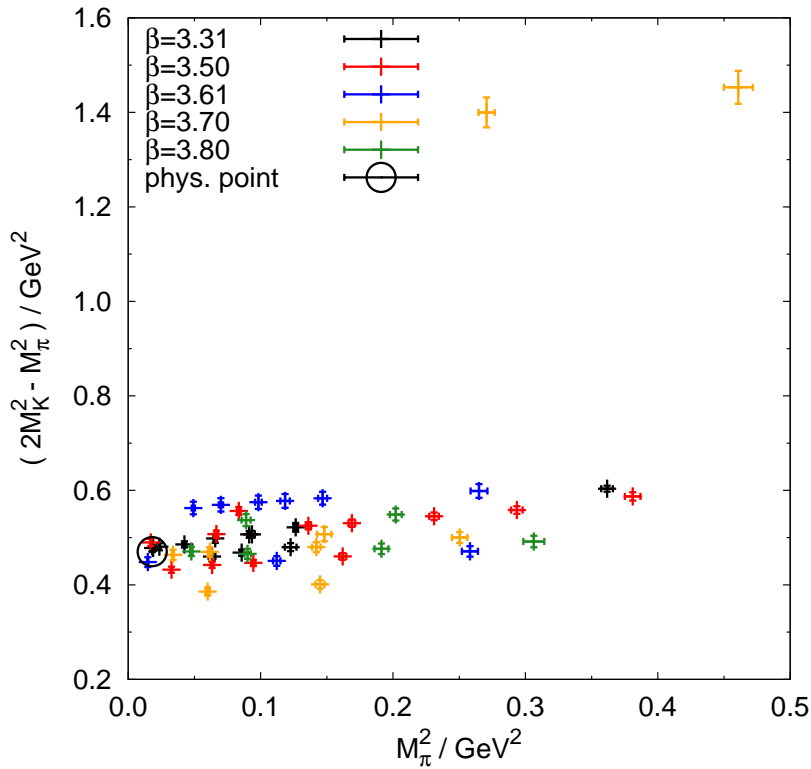}
\caption{\label{fig:ensembles}
Scatter plot of $(2M_K^2-M_\pi^2)$ vs.\ $M_\pi^2$ for all ensembles. Here the lattice scales from Tab.~\ref{tab:aInv} have been used to convert the masses from lattice units into GeV. The physical mass point, as defined in Eq.~(\ref{eq:physpoint}), is indicated by  a circle.}
\end{figure}

Our combination of actions is expected to result in cut-off effects which scale asymptotically like $\al a$, where $\al$ denotes the strong coupling constant $g^2/(4\pi)$ at the scale $a^{-1}$ (whereupon $\al$ is a logarithmic function of $a$).
In practice, cut-off effects with similar actions are often found to scale in proportion to $a^2$ over the accessible range of couplings \cite{DeGrand:1998jq,Durr:2008rw}.
This feature complicates the analysis as discussed in Sec.~\ref{sec:fitforms} below.

In Fig.~\ref{fig:ensembles} we show the combination $2M_K^2-M_\pi^2$ of squared kaon and pion masses versus the squared pion mass $M_\pi^2$ for all of our ensembles.
In this plot the $y$-axis serves as a somewhat non-linear representative of the simulated strange-quark mass $m_\mr{s}$, and the $x$-axis serves as a slightly non-linear substitute of the joint up- and down-quark mass $m_\mr{ud}$.
The non-linearity in the relation between the squared meson masses and the quark mass comes from higher orders in ChPT \cite{Gasser:1984gg}.
Note that these non-linearities do not affect the definition of the physical mass point; as long as $2\Mka^2-\Mpi^2$ is a monotonic function of $m_\mr{s}$ and $\Mpi^2$ is a monotonic function of $m_\mr{ud}$ the requirement that $(2\Mka^2-\Mpi^2)/M_\Omega^2$ and $\Mpi^2/M_\Omega^2$ would simultaneously assume their physical values leads to a unique specification of both $m_\mr{s}^\mr{phys}$ and $m_\mr{ud}^\mr{phys}$.

\begin{figure}[!tb]
\centering
\includegraphics[width=.5\textwidth]{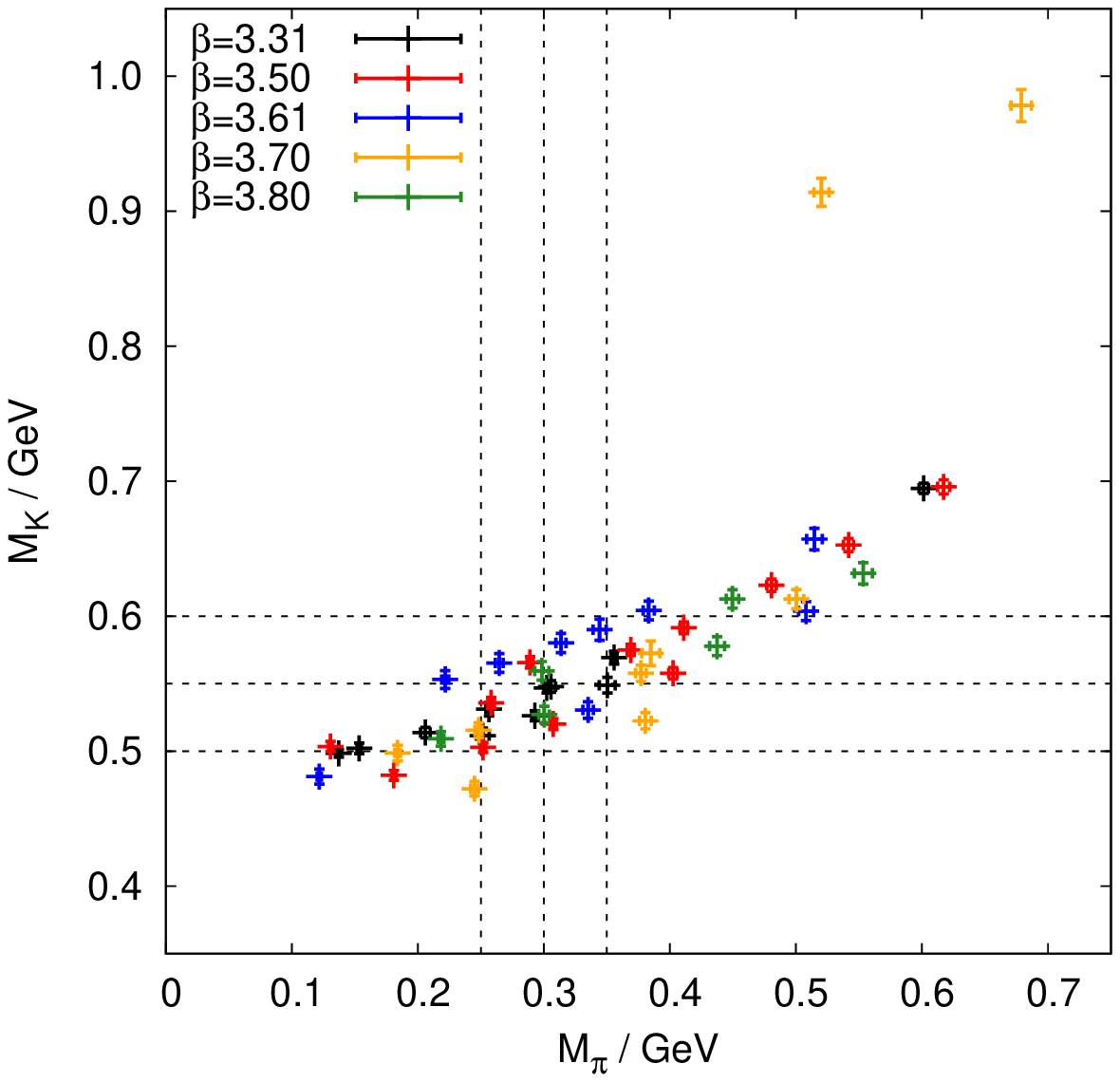}%
\includegraphics[width=.5\textwidth]{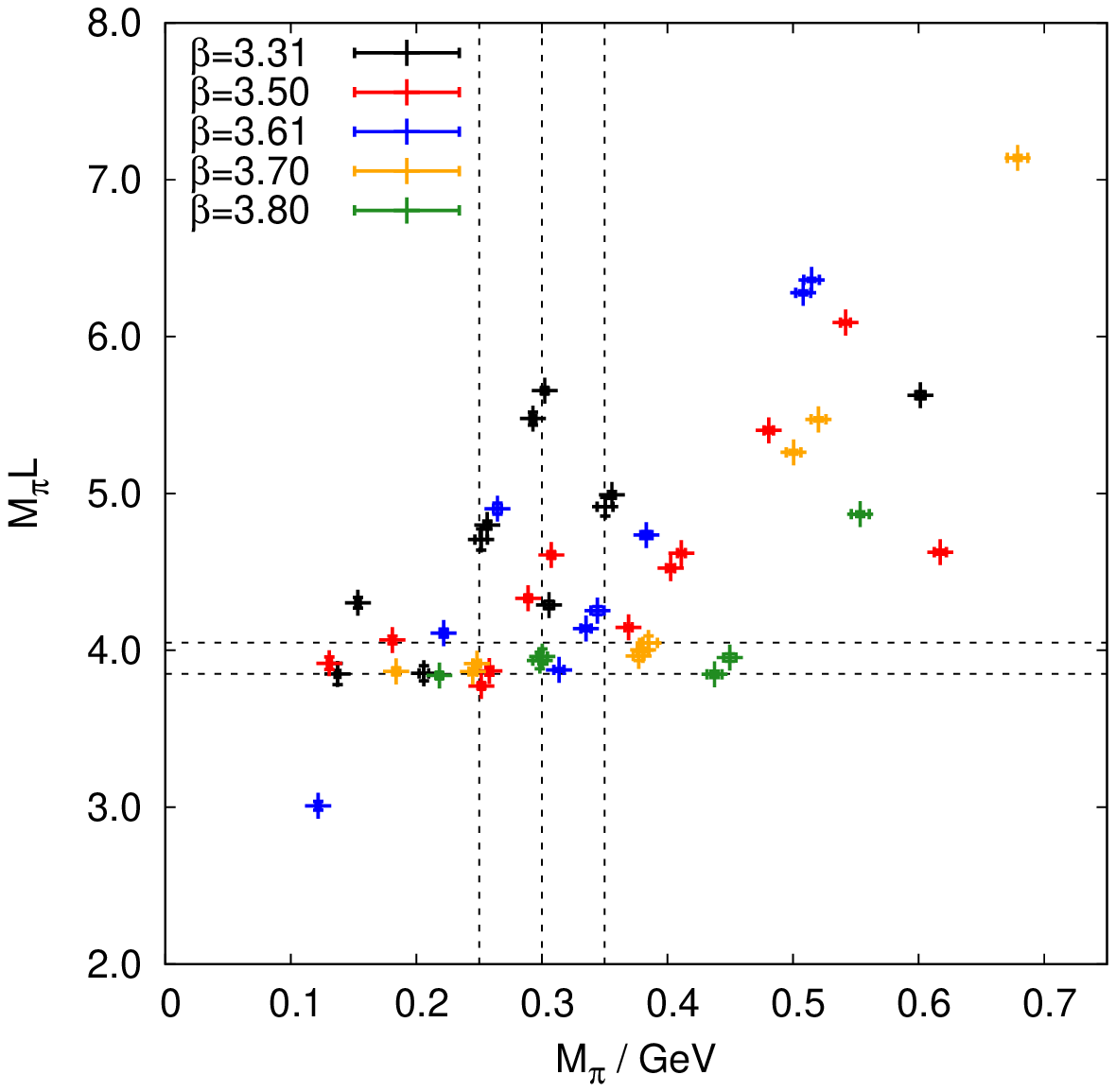}
\caption{\label{fig:fitranges}
Scatter plots of $M_K$ {\sl (left panel)} and $(M_\pi L)$ {\sl (right panel)} vs.\ $M_\pi$. Also shown are the bounds for the fit-ranges used in our analysis (see Sec.\,\ref{sec:results}). Here the lattice scales from Tab.~\ref{tab:aInv} have been used to convert the masses from lattice units into GeV.}
\end{figure}

The two panels of Fig.~\ref{fig:fitranges} show the kaon mass $M_K$ and the product $M_\pi L$, respectively, as a function of the pion mass $M_\pi$.
Also shown in these plots are the bounds for the fit ranges to be used in our analysis, which will be discussed in Sec.~\ref{sec:results} below.

Regarding the definition of the physical mass point, let us add that one should not naively use the experimental values of $M_{\pi^\pm},M_{K^\pm}$ and $M_\Omega$, since these values are affected by electromagnetic self-energies and strong isospin breakings, due to $m_\mr{u}^\mr{phys} \neq m_\mr{d}^\mr{phys}$.
In the FLAG report \cite{Aoki:2013ldr} it is discussed how such effects can be accounted for; here we follow the recommendation to use
\begin{equation}
\label{eq:physpoint}
M_\pi^\mr{phys}\;=\;134.8(0.3)\,\mr{MeV}\,,\;\;\;M_K^\mr{phys}\;=\;494.2(0.4)\,\mr{MeV}
\end{equation}
along with the PDG value $M_\Omega^\mr{PDG}=1.67245(29)\,\mr{GeV}$ from Ref.~\cite{Agashe:2014kda}.


\section{Overview of functional forms used\label{sec:fitforms}}


In this section we discuss the functional forms used for the fits of the decay-constant ratio in the global analysis to be presented in the following section.
The functional forms have to account for the two different quark-masses that come from the joint up- and down-quark mass and the separate strange-quark mass, the different lattice volumes and the different gauge couplings (i.e.\ lattice scales) used in the simulations.
By considering at least two different functional forms for each parameter-dependence (along with appropriate cuts), we will eventually be able to estimate the systematic uncertainties inherent in our determination of the decay-constant ratio at physical quark masses in the combined infinite volume and continuum limit.

Instead of using (renormalized) quark masses measured on the various ensembles or the bare quark mass input parameters ($am_\mr{ud}^\mr{bare}$, $am_\mr{s}^\mr{bare}$), we will always use the measured meson masses of the pion and the kaon, $M_\pi$ and $M_K$ respectively, to parameterize the quark mass dependence.
In leading order ChPT, the squared pion mass is proportional to the light-quark mass, $M_\pi^2|_\mr{LO}\,=\,2B_0m_\mr{ud}$, and the squared kaon mass is proportional to the sum of the two quark masses, $M_K^2|_\mr{LO}\,=\,B_0(m_\mr{ud}+m_\mr{s})$ \cite{Gasser:1984gg}.
Therefore, the combination $(2M_K^2-M_\pi^2)|_\mr{LO}=2B_0m_\mr{s}$ is in LO proportional to the strange-quark mass and can serve, beyond LO, as a non-linear substitute to parameterize the strange-quark mass dependence.
As will become evident below, our analysis does not require the absence of higher-order contributions in these relations.

In this work the dimensionless ratio $\fka/\fpi$ is considered.
Nevertheless, we must decide on the scales $a^{-1}$ of our ensembles, since we need to interpolate our data to the physical mass point and extrapolate to zero lattice spacing, $a\to0$, and to infinite box-size, $L\to\infty$.
As mentioned in the previous section, the quantity used for this purpose is the mass of the omega baryon, $M_\Omega$.
Still, there is a two-fold ambiguity regarding the scale setting procedure, and we refer to one of them as ``per ensemble scale-setting'' while the other one is referred to as ``mass-independent scale-setting''.
In both cases the lattice scale (in physical units) is obtained from
\begin{equation}
\frac{1}{a} \;=\; \frac{M_\Omega^\mr{PDG}}{(aM_\Omega)}
\end{equation}
with the PDG value of $M_\Omega$ \cite{Agashe:2014kda}.
In the former case the denominator $(aM_\Omega)$ is evaluated on each ensemble individually; this yields 47 mass-dependent scales.
In the latter case the denominator is extrapolated, for each $\be$-value, to the physical mass point (as defined in the previous section) before the relation is evaluated; this yields the five mass-independent scales listed in Tab.~\ref{tab:aInv}.
For quantities defined at the physical mass point the ``per ensemble scale-setting'' and the ``mass-independent scale-setting'' must yield consistent results (the difference is mainly how cut-off and genuine quark mass effects are split between the respective functional ans\"atze).
On the other hand for quantities whose definition involves derivatives with respect to quark masses the situation may be more tricky (see e.g.\ the discussion in Ref.~\cite{Durr:2015dna}).

Now we turn our attention to the functional forms used to parameterize the measured decay-constant ratio.
By definition the ratio $f_K/f_\pi$ has to be unity along the flavor
symmetric line $m_\mr{ud}=m_\mr{s}$ in the $(m_\mr{ud}, m_\mr{s})$
plane, where also $M_\pi=M_K$ and $f_\pi=f_K$, and in our main analysis we will only use functional forms which obey the flavor-symmetry constraint
\begin{equation}
\label{eq:flavSymConstr}
\left.\frac{f_K}{f_\pi}\right|_{m_\mr{ud}=m_\mr{s}}\;=\;1
\end{equation}
even at finite cut-off and/or finite volume.

We start our discussion with the functional forms for the mass-dependence of the ratio $f_K/f_\pi$, from which eventually the decay-constant ratio at the physical pion- and kaon-masses will be obtained.
We will either use the functional form obtained from SU(3)-ChPT up to next-to-leading order, in short NLO-SU(3) \cite{Gasser:1984gg}, or a simple polynomial expansion in the two mass-parameters.
The NLO-SU(3) expression reads
\beq
\frac{\fka}{\fpi}=
1+\frac{c_0}{2}
\left(
\frac{5}{4}\Mpi^2\log\left(\frac{\Mpi^2}{\mu^2}\right)-
\frac{1}{2}\Mka^2\log\left(\frac{\Mka^2}{\mu^2}\right)-
\frac{3}{4}\Met^2\log\left(\frac{\Met^2}{\mu^2}\right)+
c_1[\Mka^2-\Mpi^2]
\right)
\label{form_SU3}
\label{eq:SU3}
\eeq
where we use $M_\eta^2=(4M_K^2-M_\pi^2)/3$, and the fit-parameters $c_0,c_1$ relate to the QCD low-energy constants through $c_0=1/(4\pi F_0)^2$, $c_1=128\pi^2L_5(\mu)$.
Throughout, $\mu$ is the renormalization scale of the chiral effective theory.
In the event that $c_1$ or $L_5$ are quoted, FLAG recommends to do so for $\mu=770\MeV$, since this facilitates their use in phenomenology.
Still we stress that the complete Eq.~(\ref{form_SU3}) is independent of $\mu$.
For the polynomial forms, we choose to organize the expressions in the two mass parameters $M_\pi^2$ and $M_K^2-M_\pi^2$.
This has the advantage that the flavor-symmetry constraint, Eq.~(\ref{eq:flavSymConstr}), can be enforced through the second parameter alone, while the first parameter can account for the dominant light-quark mass-dependence.
In our analysis we find that we obtain reasonable fits using these three polynomial expressions
\begin{eqnarray}
\frac{f_K}{f_\pi} &=& 1\:+\:[M_K^2-M_\pi^2]\Big(c_0^\mr{3-par}\,+\,c_1^\mr{3-par}[M_K^2-M_\pi^2]\,+\,c_2^\mr{3-par}M_\pi^2\Big)
\label{eq:pol3}\\
\frac{f_K}{f_\pi} &=& 1\:+\:[M_K^2-M_\pi^2]\Big(c_0^\mr{4-par}\,+\,c_1^\mr{4-par}[M_K^2-M_\pi^2]\,+\,c_2^\mr{4-par}M_\pi^2\,+\,c_3^\mr{4-par}M_\pi^4\Big)
\label{eq:pol4}\\
\frac{f_K}{f_\pi} &=& 1\:+\:[M_K^2-M_\pi^2]\Big(c_0^\mr{6-par}\,+\,c_1^\mr{6-par}[M_K^2-M_\pi^2]\,+\,c_2^\mr{6-par}M_\pi^2\,+\,c_3^\mr{6-par}M_\pi^4 \nonumber\\
 && \phantom{1\:+\:[M_K^2-M_\pi^2]}\,+\,c_4^\mr{6-par}M_\pi^2[M_K^2-M_\pi^2]\,+\,c_5^\mr{6-par}[M_K^2-M_\pi^2]^2\Big)
\label{eq:pol6}
\end{eqnarray}
with $c^\mr{3,4,6-par}_i$ being fit-parameters.
According to the number of parameters involved these ans\"atze will be referred to as the 3-, 4-, and 6-parameter polynomial fits, respectively.

Next, we discuss how to parameterize the dependence on the gauge coupling (equivalently on the lattice spacing) in the simulations.
Eventually, this dependence defines the continuum extrapolation $a\to0$ of the decay-constant ratio.
Given the slight difference between the asymptotically guaranteed and the observed scaling pattern that was mentioned in the previous section, it would seem natural to allow for \emph{both} types of cut-off effects, i.e.\ to allow for cut-off effects proportional to $c^\mr{disc}\alpha a+d^\mr{disc}a^2$.
As noted in previous works, with such a combined ansatz both coefficients $c^\mr{disc},d^\mr{disc}$ have a tendency to be zero within errors \cite{Durr:2010vn,Durr:2010aw,Durr:2013goa,Durr:2015dna}.
To account conservatively for the presence of cut-off effects in our data we chose to invoke one or the other ansatz in consecutive form.
Hence we use an ansatz quadratic in the lattice spacing
\begin{equation}
\label{eq:aSqrDisc}
\frac{f_K}{f_\pi} \;=\; 1\:+\:\bigg(\frac{f_K}{f_\pi}(M_\pi,M_K)\,-\,1\bigg)\bigg(1\,+\,c^\mr{disc}a^2\bigg)
\end{equation}
or linear in the product $\alpha a$ of the strong coupling parameter and the lattice spacing
\begin{equation}
\label{eq:aAlphaDisc}
\frac{f_K}{f_\pi} \;=\; 1\:+\:\bigg(\frac{f_K}{f_\pi}(M_\pi,M_K)\,-\,1\bigg)\bigg(1\,+\,c^\mr{disc} \alpha a\bigg)
\end{equation}
to describe the discretization effects.
Here the term $\frac{f_K}{f_\pi}(M_\pi, M_K)$ represents the functional form used to describe the mass-dependence, i.e.\ either one of Eqs.~(\ref{eq:SU3})--(\ref{eq:pol6}).
Note that to obey the flavor-symmetry constraint, Eq.~(\ref{eq:flavSymConstr}), we assume that the discretization effects only affect the deviation of the ratio from unity.
The values of the strong coupling parameter $\alpha$ collected in Tab.~\ref{tab:aInv} represent the values of $\al$ in the $\MSbar$ scheme evaluated at the scales given in the second column.
A potential source of systematic uncertainty is that it is not clear whether one should give preference to $\Nf=4$ matching (as all of our lattice scales sitting in between $m_\mr{c}^\mr{phys}$ and $m_\mr{b}^\mr{phys}$ would suggest) or $\Nf=3$ matching (as would seem natural given that our ensembles are generated in QCD with $\Nf=2+1$).
Fortunately, it turns out that the difference is too small to matter in practice; to avoid an irrelevant near-duplication of the number of analyses we simply use the average of the $\Nf=3$ and $\Nf=4$ columns.
As an aside we mention that only the ratio of $\al$ at one $\be$-value to $\al$ at another $\be$-value matters; the whole analysis remains unchanged if the last two columns of Tab.~\ref{tab:aInv} are rescaled by a common (arbitrary) factor.
We emphasize that the ambiguity between the functional forms of Eqs.~(\ref{eq:aSqrDisc}) and (\ref{eq:aAlphaDisc}) contributes to the systematic uncertainty of our final result; if it was known for sure that the relevant cut-off effects are either $\propto \al a$ or $\propto a^2$, our final error-bar would be smaller.

Finally, let us discuss the volume dependence of the measured decay-constant ratio.
The low-energy effective theory of QCD relates the pion mass and decay constant in finite volume $\Mpi(L),\fpi(L)$ to the infinite volume counterparts $\Mpi,\fpi$ via an expansion in $\xi=\Mpi^2/(4\pi\Fpi)^2=\Mpi^2/(8\pi^2\fpi^2)$ \cite{Luscher:1985dn,Gasser:1986vb,Colangelo:2003hf,Colangelo:2005gd} and similar formulas are available for the kaon \cite{Colangelo:2005gd}.
Unfortunately, for realistic masses and box volumes this chiral expansion is found to converge rather slowly \cite{Colangelo:2003hf,Colangelo:2005gd}.
As a result of this we decided to ignore all analytical knowledge about higher-order terms and instead use the first non-trivial order version of these formulas, but with re-fitted coefficients.
This means that we use the formula
\beq
\frac{\fka(L)}{\fpi(L)}=\frac{\fka}{\fpi}
\Big(
1+c^\mr{FV}\
\Big[
\frac{5}{8}\til{g}_1(\Mpi L)-
\frac{1}{4}\til{g}_1(\Mka L)-
\frac{3}{8}\til{g}_1(\Met L)
\Big]
\Big)
\label{eq:FV}
\eeq
with the fit-parameter $c^\mr{FV}$, where FV stands for finite volume, and the definitions
\bea
\til{g}_1(z)&=&
\frac{24}{z}K_1(z)+
\frac{48}{\sqrt{2}z}K_1(\sqrt{2}z)+
\frac{32}{\sqrt{3}z}K_1(\sqrt{3}z)+
\frac{24}{2z}K_1(2z)+...
\label{def_g1til}
\\
K_1(z)&=&\sqrt{\frac{\pi}{2z}}\,
e^{-z}\,
\Big\{1+
\frac{3}{8z}-
\frac{3\cdot5}{2(8z)^2}+
\frac{3\cdot5\cdot21}{6(8z)^3}-
\frac{3\cdot5\cdot21\cdot45}{24(8z)^4}+...
\Big\}
\label{def_K1}
\eea
where $f_K/f_\pi$ represents the chosen combination of mass- and scale-dependence, as discussed previously.
More terms in the expansion of $\til{g}_1$ are available in Ref.~\cite{Colangelo:2003hf}.
To have an alternative ansatz which still has the correct asymptotic behavior for large $\Mpi L$, we also use a version where both Eq.~(\ref{def_g1til}) and Eq.~(\ref{def_K1}) are restricted to the first term.
In either case the flavor-symmetry constraint, Eq.~(\ref{eq:flavSymConstr}), is obeyed.


\section{Main results obtained in the global analysis\label{sec:results}}


In this section we discuss how we obtain the central value of $\fka/\fpi$ at the physical mass point, in the continuum and in infinite volume, as well as its statistical and systematic uncertainties, based on the ensembles presented in Sec.~\ref{sec:ensembles} and the functional forms discussed in Sec.~\ref{sec:fitforms}.

The statistical uncertainty of a given quantity will always be determined from a bootstrap-sample of size 2000 which is generated from the bootstrap-samples of the respective underlying quantities.
These bootstrap-samples will be used through every step in the fitting procedure and any further processing of the extracted parameters, in order to ensure the correct propagation of the statistical uncertainties and their correlation in every stage of our global analysis.

The systematic uncertainty, naturally, has many sources of origin.
In this work we take into account the functional forms used for the mass, continuum-limit and infinite-volume extrapolations, the ranges considered in these extrapolations, and the method used for setting the lattice scale.
Generally, we pursue the following strategy, which was also used in other studies, see e.g.~\cite{Durr:2010vn,Durr:2010aw,Durr:2013goa,Durr:2015dna}.
For each source of systematic uncertainty we consider at least two different approaches, e.g.\ two different methods of scale setting, and perform each fit once for each method.
In that way, we arrive at many different (but valid) results for $f_K/f_\pi$ at the physical mass point, in infinite volume and in the continuum.
Each such result for $f_K/f_\pi$ can be assigned a quality measure as provided by a suitable goodness-of-fit criterion obtained from the actual fit used (e.g. $\chi^2/\mr{d.o.f.}$ or $p$-value).
By considering the distribution (unweighted or weighted by a quality measure) of these results, we will be able to quote our final value as the mean of this distribution and determine the systematic uncertainty from its width (variance).
If we consider a subset of these results, e.g.\ the subset generated with one specific functional ansatz for the volume dependence, the resulting distribution will be narrower than the full distribution.
In this way we can provide an error budget, i.e.\ we can break up the full systematic error into individual contributions.

For the mass dependence, we consider four different functional forms, Eqs.~(\ref{eq:SU3}--\ref{eq:pol6}), which have two, three, four, or six fit parameters, respectively.
For the continuum extrapolation we either use the ``discretization via $a^2$'' ansatz, Eq.~(\ref{eq:aSqrDisc}), or the ``discretization via $\alpha a$'' ansatz, Eq.~(\ref{eq:aAlphaDisc}), both of which have one fit parameter.
The infinite volume extrapolation uses the functional form of Eq.~(\ref{eq:FV}), where either the full $\tilde{g}_1$ function is used or only its leading term, invoking one fit parameter in both cases.
This already leads to $4\times2\times2=16$ different combinations of functional forms used in the extrapolation of the ratio $f_K/f_\pi$, which (depending on the mass extrapolation considered) have either four, five, six, or eight fit parameters in total.
Finally, the number of fit ans\"atze is doubled due to the two methods of setting the scale (``mass-independent'' versus ``per-ensemble'', see Sec.~\ref{sec:fitforms}).
Therefore we shall consider 32 different combinations of interpolation/extrapolation and scale setting methods in our global analysis.

Another source of spread in the final distribution is the choice of ensembles used in the global analysis.
In principle, one could consider all $2^{47}=\mathcal{O}(10^{14})$
combinations that result from including or excluding any one of the 47
available ensembles.
This procedure would neither be sensible nor feasible, but the
relatively large number of ensembles does allow us to study more
carefully the systematic effects arising from the continuum and
infinite volume extrapolations as well as the interpolation to the
physical point in both pion and kaon masses.

For studying the latter two, we impose on the pion mass either no
upper bound or one of $M_\pi^\mr{max}=350\,\mr{MeV}$, $300\,\mr{MeV}$, or
$250\,\mr{MeV}$ and for the kaon $M_K^\mr{max}=600\,\mr{MeV}$, $550\,\mr{MeV}$, $500\,\mr{MeV}$.
For the volume extrapolation, we consider either no minimal bound for the parameter $(M_\pi L)$ or $(M_\pi L)^\mr{min}=3.85$ or $4.05$.
For the readers convenience these choices are indicated in the plots
of Fig.~\ref{fig:fitranges} by horizontal and vertical dashed lines.

The most important systematic effect to study is, however, the continuum
extrapolation. Since we are in the fortunate position of having
available five different lattice spacings, we could investigate the
effect of using either all of them or the finest three or four
respectively, corresponding to a minimal gauge coupling of
$\beta\geq\beta^\mr{min}=3.31$, $3.50$ and $3.61$.
As it turns out, the effect of this cut on the precision of our final
results is rather dramatic as will be
detailed below.

When performing a combination of the cuts, we only consider those combinations of fit ranges which contain at least five ensembles, since the minimal number of fit parameters in our fits is already four.
By excluding double counting of fit ranges (combination of bounds which lead to the same set of included ensembles) we arrive at 63 possible sets of fit ranges.

In conjunction with the 32 combinations of extrapolations one might expect that this number of 63 fit ranges would lead to $32\times63=2016$ different fits.
But one has to take into account that some combinations of fit ranges do not include sufficiently many ensembles to allow for fit functions with five, six, or eight parameters.
In our analysis we shall only consider ``true fits'', that is fits which have at least one degree of freedom.
Also accounting for the rare cases in which the used fitting routine would not find a unique minimum, we arrive at a total of 1368 single fits used in the global analysis.

\begin{figure}[!tb]
\centering
\includegraphics[width=.7\textwidth]{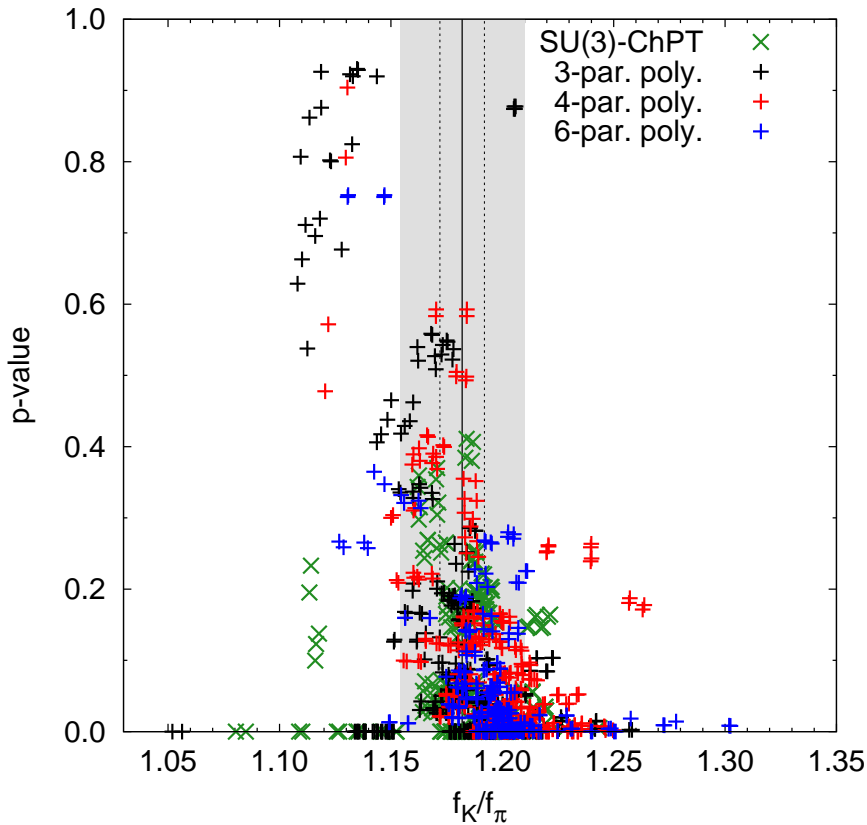}
\caption{\label{fig:scatter}
Scatter plot of the $p$-value of the fit vs.\ the physical-world ratio $f_K/f_\pi$ from each one of the 1368 single fits used in the global analysis.
The symbols distinguish fits using the ChPT or the polynomial form for the interpolation in the quark masses. Additionally, for the results using the polynomial form, the color indicates whether the 3-, 4-, or 6-parameter polynomial form has been used.
For better readability no error-bars are shown for the points in this figure.
The vertical {\sl solid} and {\sl dashed lines} and {\sl grey band} show our final result from Eq.~(\ref{eq:ratioFinal}) and its statistical and total uncertainty, respectively.}
\end{figure}

To give the reader an impression of how these fits work out in practice, we include a total of five plots.
In Fig.~\ref{fig:scatter} a scatter-plot of the $p$-value (as a goodness-of-fit measure obtained in each fit) versus the final decay-constant ratio is shown.
The grey band indicates the overall uncertainty (statistical and systematic errors added in quadrature) of our final result as specified in Eq.~(\ref{eq:ratioFinal}) below.
There is a number of low-order polynomial fits with surprisingly good $p$-values to the left of the grey band; this explains why the weighted average is smaller than the unweighted average (see below).

\begin{figure}[!tb]
\centering
\includegraphics[width=.5\textwidth]{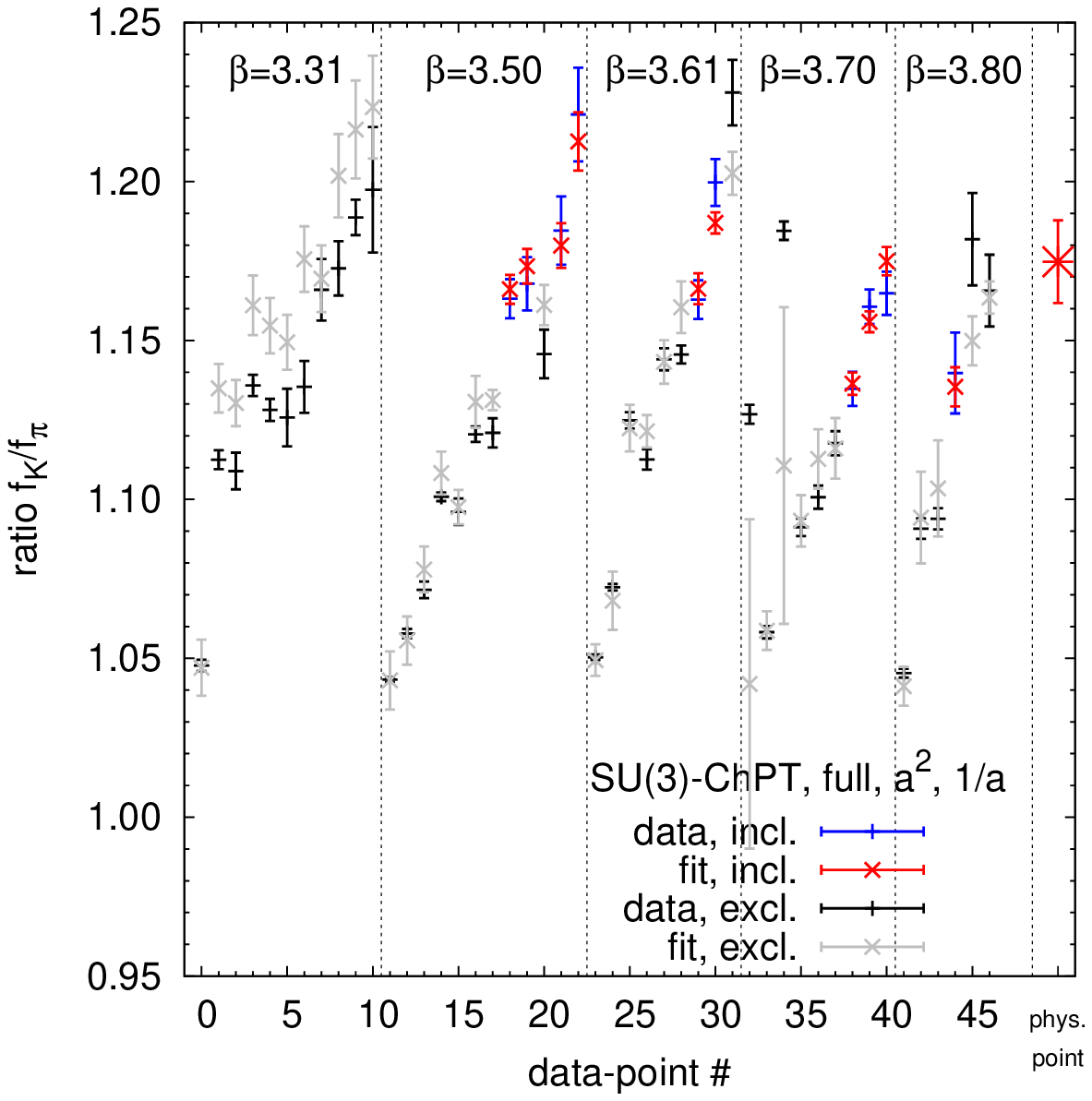}%
\includegraphics[width=.5\textwidth]{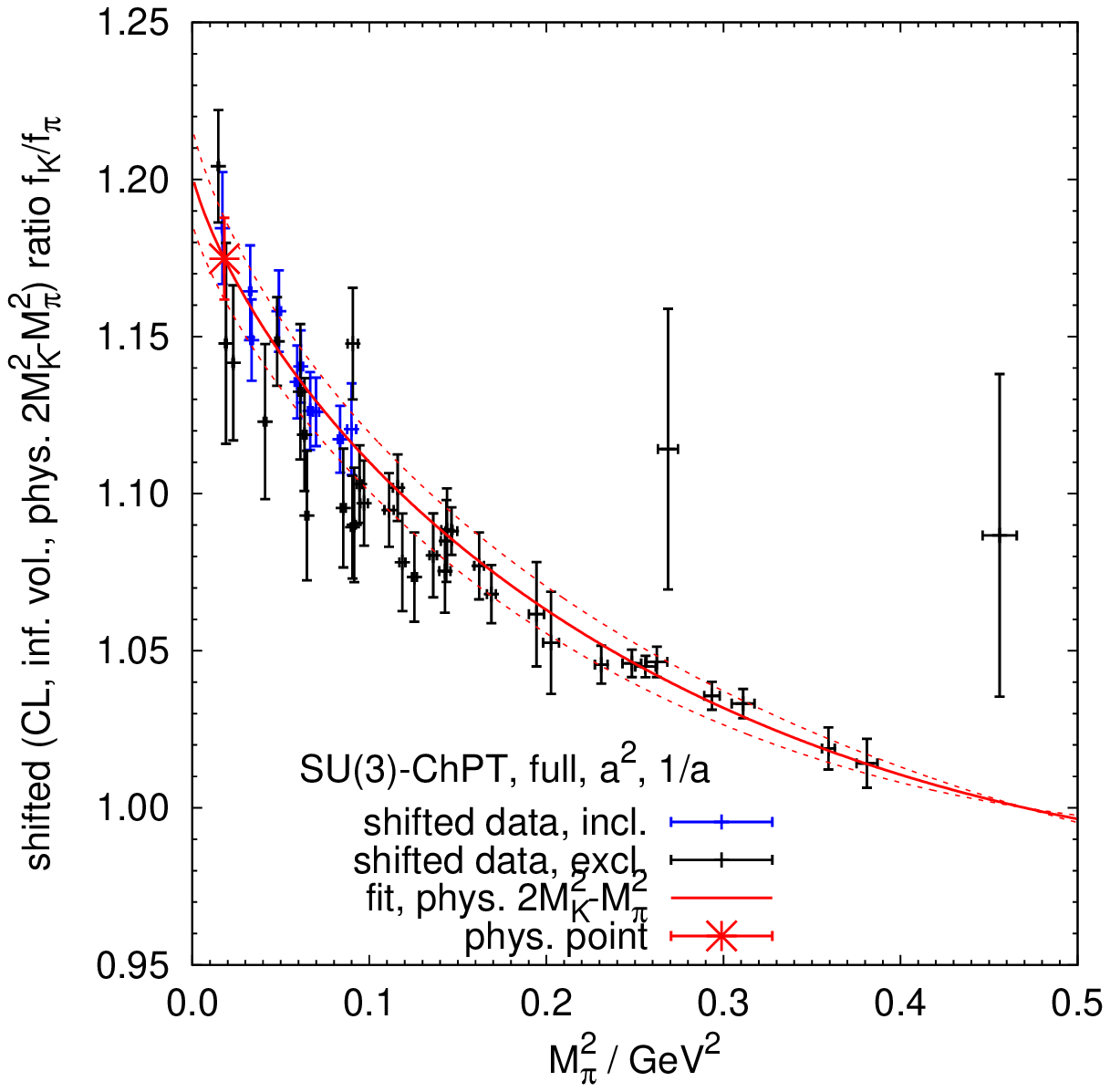}%
\caption{\label{fig:fit_SU3}
Example of a fit with SU(3)-ChPT mass-dependence, $a^2$-discretization, full FV, and mass-independent $a^{-1}$.
Included are ensembles with $M_\pi\leq300\,\mr{MeV}$, $M_K\leq600\,\mr{MeV}$, $(M_\pi L)\geq3.85$, and $\beta\geq3.50$.
The {\sl left panel} compares the value of $f_K/f_\pi$ from the fit and measured on the ensemble for each data point (labeled from 0 to 46, ordered as in Tab.~\ref{tab:ensDetails}).
The {\sl right panel} shows the dependence of the fitted $f_K/f_\pi$ on the squared pion mass $M_\pi^2$ at the physical strange-quark mass and in the infinite volume and continuum limit.
The data-points shown in this panel have been shifted according to these values/limits using the fit results.
This particular fit resulted in $\chi^2=7.6$ with $n_\mr{d.o.f.}=6$, giving a $p$-value of $0.27$.}
\end{figure}

\begin{figure}[!tb]
\centering
\includegraphics[width=.5\textwidth]{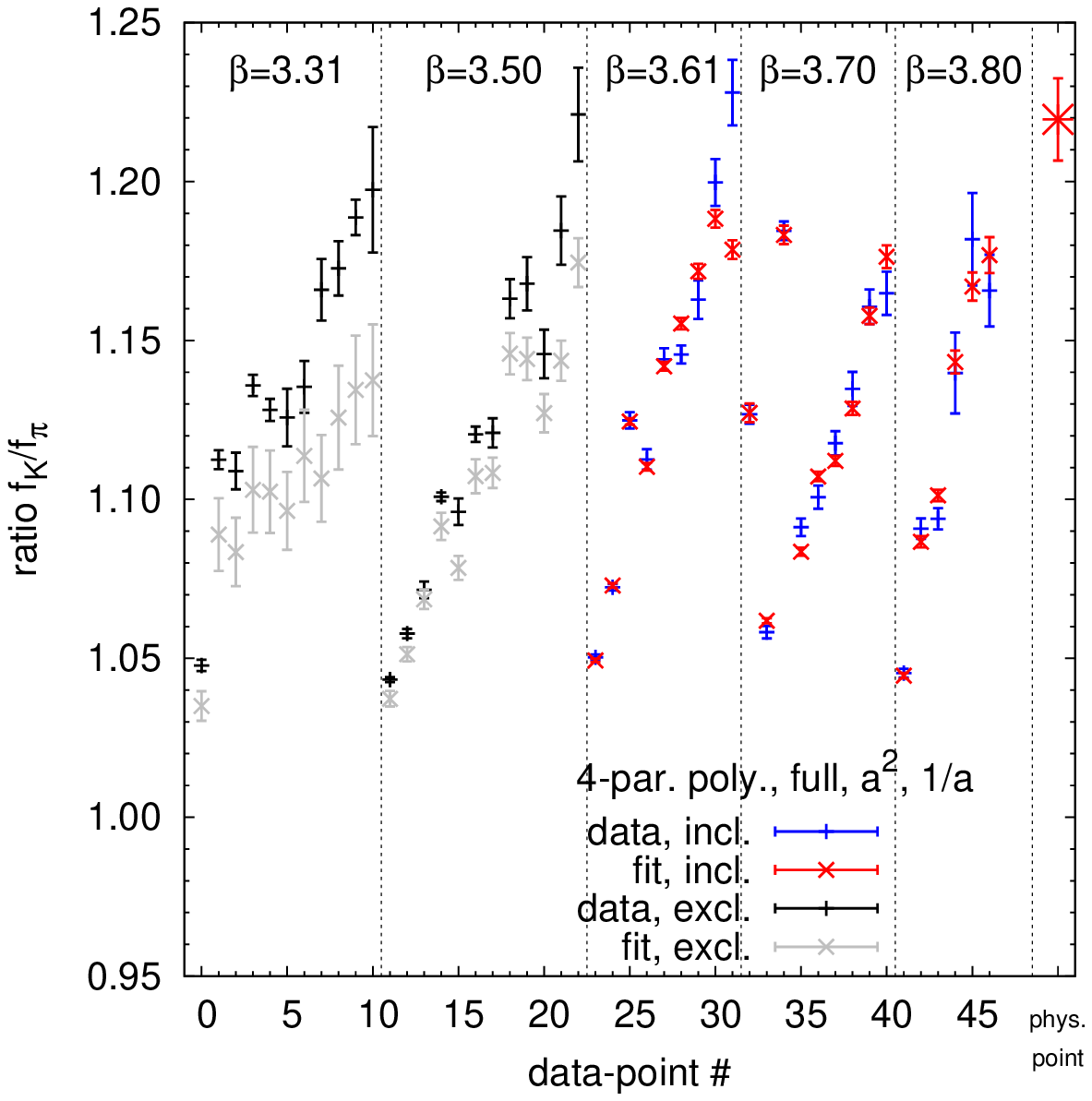}%
\includegraphics[width=.5\textwidth]{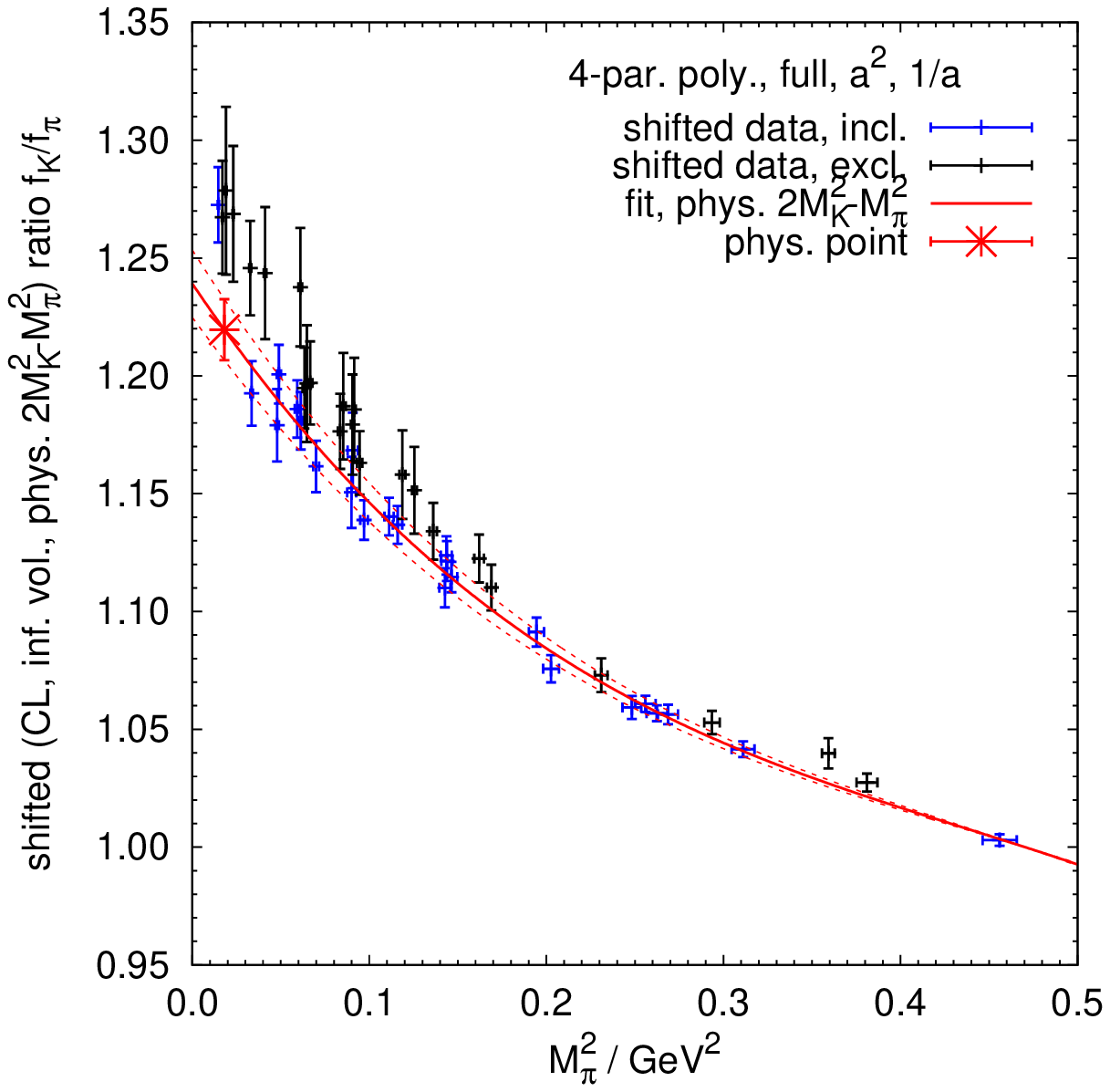}%
\caption{\label{fig:fit_pol4}
Example of a fit with 4-parameter polynomial mass-dependence, $a^2$-discretization, full FV, and mass-independent $a^{-1}$.
Included are ensembles with  $\beta\geq3.61$.
The {\sl left panel} compares the value of $f_K/f_\pi$ from the fit and measured on the ensemble for each data point (labeled from 0 to 46, ordered as in Tab.~\ref{tab:ensDetails}).
The {\sl right panel} shows the dependence of the fitted $f_K/f_\pi$ on the squared pion mass $M_\pi^2$ at the physical strange-quark mass and in the infinite volume and continuum limit.
The data-points shown in this panel have been shifted according to these values/limits using the fit results.
This particular fit resulted in $\chi^2=72.1$ with $n_\mr{d.o.f.}=18$, giving a $p$-value of $10^{-8}\approx 0$.}
\end{figure}

More details for two examples from our 1368 fits are shown in Figs.~\ref{fig:fit_SU3} and \ref{fig:fit_pol4}.
They use the SU(3)-ChPT functional form and the 4-parameter polynomial form, respectively, for the mass-dependence.
Both fits were performed with the $a^2$-discretization ansatz,  with full FV-dependence and with the mass-independent scale setting method.
The SU(3)-ChPT fit shown in Fig.~\ref{fig:fit_SU3} includes only ensembles with pion masses below $300\,\mr{MeV}$, kaon masses below $600\,\mr{MeV}$, along with $(M_\pi L)\geq3.85$, and $\beta\geq3.50$.
The 4-parameter polynomial fit shown in Fig.~\ref{fig:fit_pol4} includes all ensembles with $\beta\geq3.61$.
In general, we find that polynomial fits with suitable restrictions in the meson masses and volumes tend to give good fits (as judged by the $p$-values), but they show a wildly fluctuating behavior outside of the admitted region, especially above the selected pion mass cut.
Evidently, this behavior is neither unexpected nor does it invalidate using results extracted at the physical mass point from such a fit, since the physical meson mass region is always included in the fit range.
Perhaps the main difference between the fit shown in Fig.~\ref{fig:fit_SU3} and the one shown in Fig.~\ref{fig:fit_pol4} is that the former one (with good $\chi^2$) describes the data outside of the fit interval (represented by black symbols in the right panel) quite well, whereas the latter one (with poor $\chi^2$) misses the data outside of its fit range quite visibly.

\begin{table}[!tb]
\centering
\begin{tabular}{lccc}
\hline\hline
fit type                                    &      flat     &   $p$-value   & valid fits \\
\hline\hline
all                                         & 1.191(08)(24) & 1.173(11)(29) &    1368    \\
\hline
$\Mpi^\mr{max}=250\MeV$                     & 1.194(18)(29) & 1.180(21)(43) & 104\\
$\Mpi^\mr{max}=300\MeV$                     & 1.184(12)(22) & 1.168(15)(28) & 222\\
$\Mpi^\mr{max}=350\MeV$                     & 1.196(09)(23) & 1.177(11)(21) & 380\\
$\Mpi^\mr{max}=\infty$                      & 1.189(07)(24) & 1.171(10)(25) & 662\\
\hline
$\Mka^\mr{max}=500\MeV$                     & 1.190(27)(12) & 1.197(33)(11) & 12\\
$\Mka^\mr{max}=550\MeV$                     & 1.182(11)(18) & 1.174(12)(26) & 396\\
$\Mka^\mr{max}=600\MeV$                     & 1.198(09)(23) & 1.168(13)(32) & 680\\
$\Mka^\mr{max}=\infty$                      & 1.186(06)(28) & 1.188(13)(23) & 280\\
\hline
$(\Mpi L)^\mr{min}=0$                       & 1.191(10)(21) & 1.169(15)(33) & 623\\
$(\Mpi L)^\mr{min}=3.85$                    & 1.191(10)(24) & 1.175(14)(23) & 497\\
$(\Mpi L)^\mr{min}=4.05$                    & 1.188(10)(31) & 1.179(16)(28) & 248\\
\hline
$\be^\mr{min}=3.31$                         & 1.194(08)(17) & 1.193(12)(20) & 682\\
$\be^\mr{min}=3.50$                         & 1.183(09)(17) & 1.168(13)(22) & 450\\
$\be^\mr{min}=3.61$                         & 1.195(14)(42) & 1.135(23)(25) & 236\\
\hline
scale setting mass-independent              & 1.188(08)(27) & 1.168(12)(32) & 719\\
scale setting per-ensemble                  & 1.194(08)(20) & 1.179(12)(24) & 649\\
\hline
discretization via $a^2$-term               & 1.189(07)(24) & 1.169(12)(31) & 700\\
discretization via $\al a$-term             & 1.193(09)(24) & 1.178(11)(24) & 668\\
\hline
finite volume via full $\til{g}_1$ function & 1.191(08)(24) & 1.172(11)(29) & 688\\
finite volume via leading terms             & 1.191(08)(24) & 1.174(11)(28) & 680\\
\hline
quark mass function via ChPT                & 1.189(07)(21) & 1.183(10)(19) & 341\\
quark mass function via 3-par.\ poly.\      & 1.181(08)(26) & 1.159(14)(30) & 382\\
quark mass function for 4-par.\ poly.\      & 1.198(09)(21) & 1.183(13)(27) & 367\\
quark mass function for 6-par.\ poly.\      & 1.197(09)(22) & 1.177(11)(27) & 278\\
\hline\hline
\end{tabular}
\caption{\label{tab:fitresults}
Comparison of the unweighted (``flat'') and $p$-value weighted results from the complete set of fits (top line) with the various subsets (according to $\Mpi^\mr{max}$, $\Mka^\mr{max}$, $(\Mpi L)^\mr{min}$, $\be^\mr{min}$, the scale setting method, the type of discretization terms, the type of finite volume terms, and the kind of quark mass dependence used) which allows for a breakup of the systematic uncertainty into various sources (see text for details). The first error is statistical and the second one is systematic. The last column gives the number of single fits which match these criteria.}
\end{table}

The final step is to perform an average of our 1368 fits, each of which has a central value and a statistical uncertainty.
This average may be performed with a uniform weight for all fits or it may be performed in a weighted manner, and a similar statement holds for the standard deviation of the 1368 fitted values, which we use as a measure of the systematic uncertainty of the aggregate $\fka/\fpi$.
Regarding the weights, it is natural to consider weights which derive from some goodness-of-fit number of the individual fits.
One option is to use a weight proportional to the $p$-value of each fit.
In Tab.~\ref{tab:fitresults} we collect the various averages that result from our 1368 valid analyses with a flat weight in the left column and with the $p$-value weight in the right column.
In addition, the resulting aggregate values (along with their statistical and systematic uncertainties) are shown for all the data cuts considered and for all functional forms employed; this amounts to a break-up of the overall uncertainty into its various contributions.
Instead of using the $p$-value as a weight, we tried alternatives such as the weights based on the Akaike information criterion \cite{Akaike:1974} or based on $\chi^2/n_\mr{d.o.f.}$.
Basically, we find that using these weights does not change the mean values and error estimates significantly compared to the results obtained with the $p$-value, and this is why we refrain from including them in the table.

%
\begin{figure}[!tb]
\centering
\includegraphics[width=.45\textwidth]{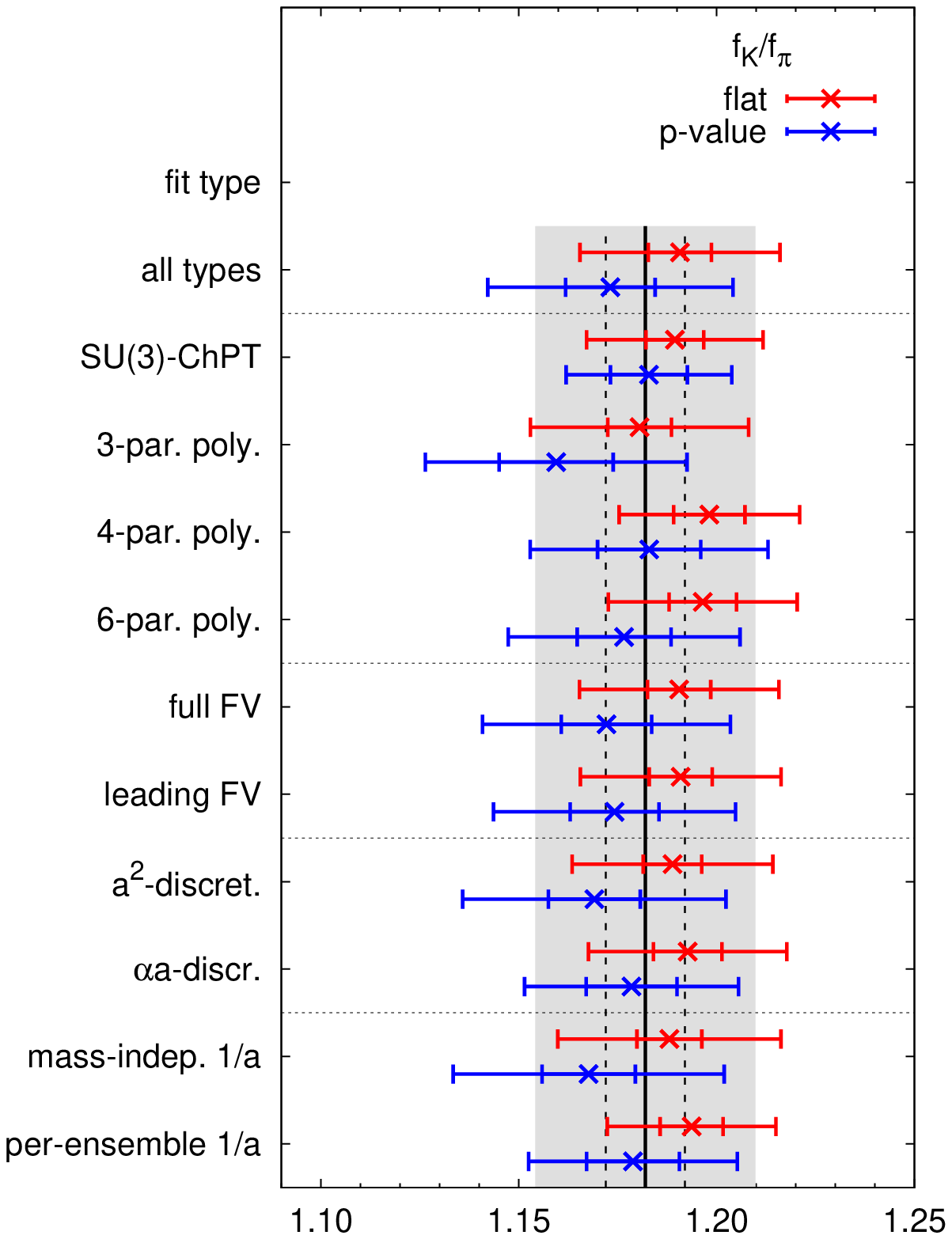}%
\hspace*{.05\textwidth}%
\includegraphics[width=.45\textwidth]{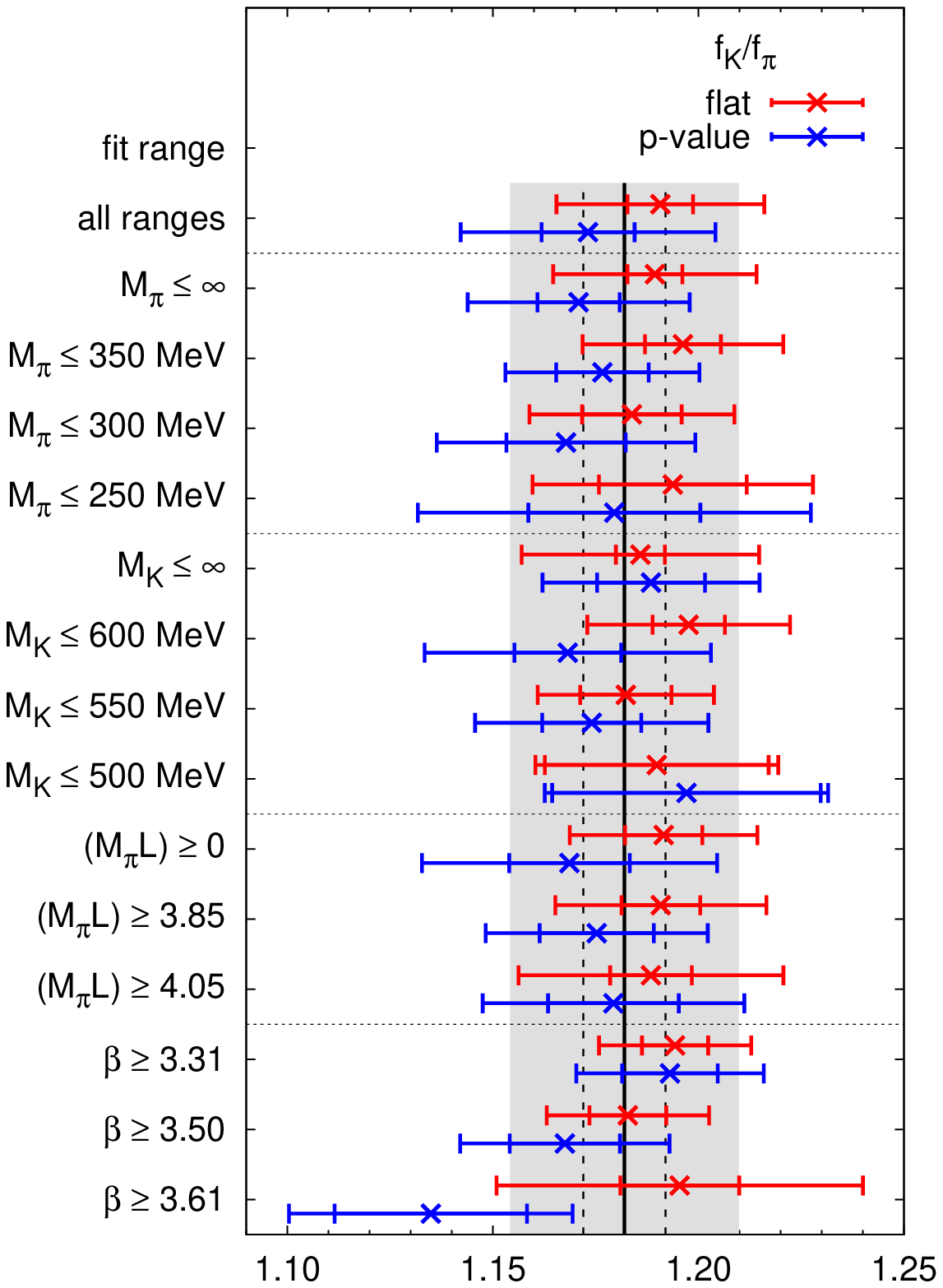}
\caption{\label{fig:fitresults}
Overview of the results from the global analysis with different fit types {\sl (left)} and different fit ranges {\sl (right)}.
Shown are the value and the statistical and combined statistical and systematic errors, see Tab.~\ref{tab:fitresults} for details.
The {\sl vertical black solid and dashed lines} and the {\sl grey shaded band} show the central value, the statistical and combined error, respectively, of our final result for $f_K/f_\pi$ in 2+1 flavor QCD as given in Eq.~(\ref{eq:ratioFinal}).}
\end{figure}

A graphical summary of the content of Tab.~\ref{tab:fitresults} is shown in Fig.~\ref{fig:fitresults}.
The black vertical line indicates our central value as given in Eq.~(\ref{eq:ratioFinal}) below, the dashed set of vertical lines shows the statistical uncertainty, and the grey band displays the combined uncertainty as specified in that equation.
For each fit ansatz (left panel) and cut on the data (right panel) the respective entry in Tab.~\ref{tab:fitresults} is shown, enabling one to spot trends (e.g.\ in $\be^\mr{min}$) more easily.

To summarize Tab.~\ref{tab:fitresults}, we note that from the unweighted and $p$-value weighted distributions we obtain from all fits the final results
\bea
\label{eq:ratioFlat}
\fka/\fpi|_\mr{flat}\,    &=&1.191(08)_\mr{stat}(24)_\mr{syst}=1.191(25)_\mr{comb}\,,
\\
\label{eq:ratioPval}
\fka/\fpi|_{p\mr{-value}} &=&1.173(11)_\mr{stat}(29)_\mr{syst}=1.173(31)_\mr{comb}\,,
\eea
where the combined error was obtained by adding the statistical and systematic errors in quadrature.
The only question left open is to which one of the two weighting strategies preference should be given.
In the end, we decided to take a straight average, resulting in
\beq
\label{eq:ratioFinal}
\fka/\fpi=1.182(10)_\mr{stat}(26)_\mr{syst}=1.182(28)_\mr{comb}
\eeq
as our final result for the decay-constant ratio in $\Nf=2+1$ QCD at the physical mass point in the combined infinite volume and continuum limit.

The analysis just presented is in many ways very conservative. Most
importantly, we estimated the systematic error of the continuum
extrapolation by both varying between $\alpha a$ and $a^2$ terms and
optionally throwing away ensembles at our two coarsest lattice
spacings. As it turns out, the coefficient $c^\mr{disc}$ of the
continuum extrapolations in
 Eqs.~(\ref{eq:aSqrDisc}), (\ref{eq:aAlphaDisc}) is typically zero within our
 statistical error if we disregard ensembles at our coarsest lattice
 spacing, i.e. $\beta_\mr{min}=\{3.5,3.61\}$, which could be an indication
 that we overestimate our systematical error.

In a similar fashion, the flavor symmetry constraint
(\ref{eq:flavSymConstr}) precludes us from using fit functions which
effectively describe the behaviour of $f_K/f_\pi$ around the physical
point but fail in the SU(3) symmetric case. Again, this procedure
results in a very conservative estimate of the pertaining systematic
error, since fit functions that describe data accurately in a range
from the physical point to the SU(3) symmetric line typically need a
larger number of fit parameters.

To address these issues, we performed two complete supplementary
analyses. In these analyses, the continuum extrapolation includes the
standard $\alpha a$ and $a^2$ choices, as well as an additional option
of having no term for cut-off effects at all (referred to as ``const''
below).

To model the quark mass dependence around the physical
point, we included an NLO SU(2)-ChPT fit function, as well as Taylor
and Pade ans\"atze in the variables $\Mpi^2$ and $2\Mka^2-\Mpi^2$, as
detailed in Ref.~\cite{Durr:2010hr}. These ans\"atze do not obey the
flavor symmetry constraint (\ref{eq:flavSymConstr}). Additionally, we
put an upper limit of four on the number of parameters for our chiral fit
functions, which eliminates the 6-parameter flavor-breaking
fit-function (\ref{eq:pol6}). The values for the cuts in the pion mass
were adjusted to $275$, $350$, and $400\MeV$, the values for the cut
in $(\Mpi L)$ were unchanged, as was the scale setting. Last but not
least, the minimum number of degrees of freedom of a fit to be
included was four (as opposed to one in the main analysis).

With these modifications, we obtain
$f_K/f_\pi=1.198(08)_\mr{stat}(32)_\mr{syst}$ with flat weights and
$f_K/f_\pi=1.186(08)_\mr{stat}(11)_\mr{syst}$ with $p$-value
weighting, which has to be compared to the main analysis results in
Eqs.~(\ref{eq:ratioFlat}), (\ref{eq:ratioPval}). This constitutes our
first supplementary analysis.

In the second one of our supplementary analyses we additionally eliminated
procedures in which the cutoff-dependence might be
overfitted. Specifically, we introduced a linkage between the cut in
$\beta$ and the type of cut-off ansatz: the two non-trivial cut-off
term options ($a^2$ and $\alpha a$) are only used with the trivial
$\beta$-cut ($\beta_{\rm min}=3.31$), that is out of the
$3\!\times\!3$ combined options for treating discretization effects
only five are used. With flat weight this analysis results in
$f_K/f_\pi=1.192(07)_\mr{stat}(15)_\mr{syst}$, which again can be
compared to Eq.~(\ref{eq:ratioFlat}) from our main analysis.
With $p$-value
weighting, we obtain 
\begin{equation}
\label{eq:Supp}
f_K/f_\pi=1.188(09)_\mr{stat}(09)_\mr{syst}
\end{equation}
which can be compared to Eq.~(\ref{eq:ratioPval}) as well as to
$f_K/f_\pi=1.192(07)_\mr{stat}(06)_\mr{syst}$, which was the final
result obtained in Ref.~\cite{Durr:2010hr}.


\section{Discussion\label{sec:discussion}}


The decay-constant ratio in Eq.~(\ref{eq:ratioFinal}) represents our final result for QCD with two degenerate flavors taken at the average mass of the up- and down-quarks in the real world and a single flavor taken at the physical strange-quark mass.
Its systematic error includes all sources of theoretical uncertainty in that theory.
However, the quantity we are after is $f_{K^\pm}/f_{\pi^\pm}$ in the real world, i.e.\ in QCD with six non-degenerate flavors, each of which to be taken at its own physical mass.
This change will bring a shift of the central value, as well as an increase in the systematic uncertainty, and our goal is to discuss these effects.

Regarding the influence of those quark flavors which were ignored in Eq.~(\ref{eq:ratioFinal}) it is clear that the dominant unquenching effect comes from the lightest flavor ignored, i.e.\ the charm quark.
Since the QCD functional determinant is quadratic in the masses of the quarks, naive reasoning would suggest that unquenching effects from charm quark loops are suppressed, relative to those from strange quarks, by a factor $(m_\mr{c}/m_\mr{s})^2\simeq140$.
Since already the presence of strange quark loops in the sea seems to affect $\fka/\fpi$ very little \cite{Aoki:2013ldr}, we take it for granted that the quenching of the heavier flavors introduces an uncertainty which is small in comparison to the one which is declared in Eq.~(\ref{eq:ratioFinal}) and can thus be considered negligible in what follows.

The correction needed to undo the isospin symmetry limit in Eq.~(\ref{eq:ratioFinal}) is also small compared to the uncertainty in that equation.
However, the issue has been analyzed in ChPT \cite{Gasser:1984gg,Cirigliano:2011tm}, relating the object of interest, $f_{K^\pm}/f_{\pi^\pm}$, to the isospin symmetric quantity $\fka/\fpi$ through
\beq
\frac{f_{K^\pm}}{f_{\pi^\pm}}=
\frac{\fka}{\fpi}
\sqrt{1+\de_\mr{SU(2)}}
\eeq
but the information on $\de_\mr{SU(2)}$ in the literature is not very conclusive.
The original work suggests $\de_\mr{SU(2)}=-0.0043(12)$ \cite{Cirigliano:2011tm} and FLAG finds, from a re-analysis of several $\Nf=2+1$ computations, a very similar value \cite{Aoki:2013ldr}.
On the other hand, a dedicated study in $\Nf=2$ QCD ends up finding $\de_\mr{SU(2)}=-0.0078(7)$ \cite{deDivitiis:2011eh}.
In this situation we opt for $\de_\mr{SU(2)}=-0.0061(61)_\mr{syst}$ with a generous 100\% error.
Upon combining Eq.~(\ref{eq:ratioFinal}) with this estimate of SU(2) breaking effects, we find
\beq
\label{eq:ratioPheno}
f_{K^\pm}/f_{\pi^\pm}=1.178(10)_\mr{stat}(26)_\mr{syst}=1.178(28)_\mr{comb}
\eeq
which is the quantity to be used in the final phenomenological
analysis.
Combining it instead with the result (\ref{eq:Supp}) of our alternative analysis we find
\beq
f_{K^\pm}/f_{\pi^\pm}=1.184(9)_\mr{stat}(11)_\mr{syst}=1.184(14)_\mr{comb}.
\eeq

The next ingredient is an evaluated form of the original Marciano relation, Eq.~(\ref{eq:marciano}), with all uncertainties adequately propagated.
In the literature we find \cite{Moulson:2014cra,Rosner:2015wva}
\beq
\label{eq:evaluated}
\frac{V_\mr{us}}{V_\mr{ud}}\frac{f_{K^\pm}}{f_{\pi^\pm}}=0.27599(29)(24)=0.27599(38)_\mr{exp}
\eeq
which, when combined with our result (\ref{eq:ratioPheno}), yields
\beq
\label{eq:ratioCKM}
\frac{V_\mr{us}}{V_\mr{ud}}=0.2343(20)_\mr{stat}(52)_\mr{syst}(03)_\mr{exp}=0.2343(55)_\mr{comb}
\eeq
where ``stat'' and ``syst'' refer to our statistical and systematic uncertainties, respectively, while ``exp'' refers to the combined uncertainty of Eq.~(\ref{eq:evaluated}), which is comparatively small.

With this ratio of CKM matrix elements in hand, and given that $|V_\mr{ub}|=4.12(37)(06)\cdot10^{-3}$ \cite{Rosner:2015wva} is small on the scale of our uncertainties, one can proceed in two ways.
One option is to assume the unitarity inherent in the CKM paradigm.
In this case the square of Eq.~(\ref{eq:ratioCKM}) is augmented with the first-row unitarity relation in the form $V_\mr{ud}^2\cdot(1+V_\mr{us}^2/V_\mr{ud}^2)=0.999983(3)$ to yield
\beq
\label{eq:Vud_unitarity}
V_\mr{ud}=0.9736(04)_\mr{stat}(11)_\mr{syst}(01)_\mr{exp}=0.9736(12)_\mr{comb}
\;,
\eeq
while the inverse square of Eq.~(\ref{eq:ratioCKM}) is combined with the first-row unitarity relation in the form $V_\mr{us}^2\cdot(1+V_\mr{ud}^2/V_\mr{us}^2)=0.999983(3)$ to yield
\beq
\label{eq:Vus_unitarity}
V_\mr{us}=0.2281(18)_\mr{stat}(48)_\mr{syst}(03)_\mr{exp}=0.2281(51)_\mr{comb}
\;.
\eeq
The other option is to multiply the very precise result from super-allowed nuclear beta decays, $V_\mr{ud}=0.97417(21)_\mr{nuc}$ by Hardy and Towner \cite{Hardy:2014qxa}, with our result Eq.~(\ref{eq:ratioCKM}) to give
\beq
V_\mr{us}=0.2282(19)_\mr{stat}(51)_\mr{syst}(03)_\mr{exp+nuc}=0.2282(54)_\mr{comb}
\;,
\eeq
and in the same go one may form the product $V_\mr{ud}^2\cdot(1+V_\mr{us}^2/V_\mr{ud}^2)$ with the Hardy-Towner value as the first factor and our result Eq.~(\ref{eq:ratioCKM}) as the second factor and add $|V_\mr{ub}|^2$ to find
\beq
\label{eq:check}
V_\mr{ud}^2+V_\mr{us}^2+|V_\mr{ub}|^2-1=0.0011(09)_\mr{stat}(23)_\mr{syst}(05)_\mr{exp+nuc}=0.0011(25)_\mr{comb}
\eeq
which indicates that the unitarity relation is well obeyed, within errors.


\section{Summary\label{sec:sumary}}


In this paper we have presented a lattice computation of $\fka/\fpi$ in 2+1 flavor QCD, i.e.\ in the isospin limit with the two light quarks taken at the average of the physical up- and down-quark masses, while the non-degenerate last flavor is taken at the physical value of the strange-quark mass.
The final value in Eq.~(\ref{eq:ratioFinal}) gives the combined continuum and infinite volume limit, and it accounts for all sources of systematic uncertainty in that theory.

In the next step we have used external information to account for the breaking of isospin symmetry in the real world.
Fortunately, the correction is very small, and our result for $f_{K^\pm}/f_{\pi^\pm}$ as given in Eq.~(\ref{eq:ratioPheno}) has the same error-bar as its isospin symmetric counterpart.
Still, when comparing to the literature we find that our uncertainty is fairly large.
The MILC value $f_{K^\pm}/f_{\pi^\pm}=1.1947(26)(37)$ \cite{Bazavov:2013vwa},
the HPQCD value $f_{K^\pm}/f_{\pi^\pm}=1.1916(15)(15)$ \cite{Dowdall:2013rya},
the Fermilab Lattice/MILC value $f_{K^\pm}/f_{\pi^\pm}=1.1956(10)(^{+26}_{-18})$ \cite{Bazavov:2014wgs}
and the ETMC value $f_{K^\pm}/f_{\pi^\pm}=1.184(12)(11)$ \cite{Carrasco:2014poa}
all stem from simulations with $\Nf=2+1+1$ dynamical flavors, and concern the charged decay-constant ratio.
The RBC/UKQCD value $\fka/\fpi=1.1945(45)$ \cite{Blum:2014tka} comes from simulations with $\Nf=2+1$ flavors of domain-wall fermions, and requires the same (small) isospin breaking correction that we have applied to our result.
What catches our attention is that the overall uncertainties quoted by ETMC and in our Eq.~(\ref{eq:ratioPheno}) are roughly an order of magnitude larger than the overall uncertainties obained with staggered and domain-wall fermions, but we are unaware of a convincing explanation why this would naturally be so.

In a last step we have explored the implications of our result for the charged decay-constant ratio on the first row of the CKM matrix.
By invoking only experimental information we find Eq.~(\ref{eq:ratioCKM}).
If we assume that the CKM matrix is unitary, we find the individual elements Eq.~(\ref{eq:Vud_unitarity}) and Eq.~(\ref{eq:Vus_unitarity}), respectively.
While not very precise, they are at least consistent with the averages given in Refs.~\cite{Aoki:2013ldr,Rosner:2015wva}.
Alternatively, we may take the Hardy-Towner value of $V_\mr{ud}$ as an additional input.
In this case we can test whether our result is consistent with the CKM matrix being unitary, and Eq.~(\ref{eq:check}) tells us that, within errors, this is the case.

\subsection*{Acknowledgments}

Computations were carried out on the BG/Q supercomputer JUQUEEN at Forschungszentrum J\"ulich through a NIC grant, on Turing at IDRIS, France, under GENCI-IDRIS grant 52275, and on a local cluster at the University of Wuppertal.
This work was supported by the DFG through the SFB/TRR 55 ``Hadron Physics from Lattice QCD'', by the EU Framework Programme 7 grant (FP7/2007-2013)/ERC No 208740, by the OTKA grant OTKA-NF-104034, and by the projects OCEVU Labex (ANR-11-LABX-0060) and A*MIDEX (ANR-11-IDEX-0001-02).
In addition, E.E.S.\ acknowledges support from the EU grant PIRG07-GA-2010-268367.

\bigskip

\newpage
\appendix\section*{Appendix: Ensemble details}


In this appendix we arrange some details on the ensembles used in this work.
To simulate the $N_f=2+1$ flavors with the clover-improved fermion action with 2-HEX smearing and the Symanzik-improved gauge action, the Hybrid-Monte-Carlo (HMC) algorithm has been used together with the rational approximation variety (RHMC) to represent the single strange quark flavor.
More details on the implementation of these algorithms can be found in Ref.~\cite{Durr:2010aw}.

\setlength{\LTcapwidth}{\textwidth}
\begin{longtable}{c*{2}{r}*{5}{l}}
%
\hline\hline%
\multicolumn{1}{c}{$\beta$} & \multicolumn{1}{c}{$am_\mr{ud}^\mr{bare}$} & \multicolumn{1}{c}{$am_\mr{s}^\mr{bare}$} & \multicolumn{1}{c}{$L/a$} & \multicolumn{1}{c}{$aM_\pi$} & \multicolumn{1}{c}{$aM_K$} & \multicolumn{1}{c}{$aM_\Omega$} & \multicolumn{1}{c}{$f_K/f_\pi$} \\\hline\hline
\endfirsthead
\hline\hline%
\multicolumn{1}{c}{$\beta$} & \multicolumn{1}{c}{$am_\mr{ud}^\mr{bare}$} & \multicolumn{1}{c}{$am_\mr{s}^\mr{bare}$} & \multicolumn{1}{c}{$L/a$} & \multicolumn{1}{c}{$aM_\pi$} & \multicolumn{1}{c}{$aM_K$} & \multicolumn{1}{c}{$aM_\Omega$} & \multicolumn{1}{c}{$f_K/f_\pi$} \\\hline\hline
\multicolumn{8}{c}{continued from previous page}\\
\endhead
\multicolumn{8}{c}{to be continued on next page}\\\hline\hline
\endfoot
\hline\hline\\
\caption{\label{tab:ensDetails}
Simulated ensembles and measured $aM_\pi$, $aM_K$, $aM_\Omega$, and $f_K/f_\pi$.}
\endlastfoot
3.31 & -0.07000 & -0.0400 & 16 & 0.35262(104) & 0.40757(96) & 1.1364(189) & 1.0478(18) \\
     & -0.09000 & -0.0400 & 24 & 0.20838(76)  & 0.33265(64) & 1.0408(90)  & 1.1125(30) \\
     & -0.09000 & -0.0440 & 24 & 0.20266(121) & 0.31998(95) & 1.0327(144) & 1.1089(58) \\
     & -0.09300 & -0.0400 & 24 & 0.17789(57)  & 0.32029(40) & 1.0126(47)  & 1.1359(33) \\
     & -0.09300 & -0.0400 & 32 & 0.17697(51)  & 0.32001(42) & 1.0212(52)  & 1.1281(35) \\
     & -0.09300 & -0.0440 & 32 & 0.17182(112) & 0.30812(83) & 0.9912(107) & 1.1258(90) \\
     & -0.09530 & -0.0400 & 32 & 0.14967(74)  & 0.30985(65) & 1.0141(81)  & 1.1354(82) \\
     & -0.09530 & -0.0440 & 32 & 0.14534(120) & 0.29809(75) & 1.0014(114) & 1.1660(97) \\
     & -0.09756 & -0.0400 & 32 & 0.11928(98)  & 0.30052(59) & 0.9898(60)  & 1.1727(85) \\
     & -0.09900 & -0.0400 & 48 & 0.08950(69)  & 0.29352(35) & 0.9861(36)  & 1.1887(56) \\
     & -0.09933 & -0.0400 & 48 & 0.08111(107) & 0.29257(87) & 0.9918(110) & 1.1974(198)\\ \hline
3.50 & -0.02500 & -0.0060 & 16 & 0.28911(79)  & 0.32587(72) & 0.8983(165) & 1.0433(08) \\
     & -0.03100 & -0.0060 & 24 & 0.25373(45)  & 0.30566(43) & 0.8654(85)  & 1.0578(15) \\
     & -0.03600 & -0.0060 & 24 & 0.22506(71)  & 0.29175(62) & 0.8553(105) & 1.0716(26) \\
     & -0.04100 & -0.0060 & 24 & 0.19249(46)  & 0.27697(41) & 0.8332(37)  & 1.1008(14) \\
     & -0.04100 & -0.0120 & 24 & 0.18851(77)  & 0.26121(71) & 0.8058(146) & 1.0961(42) \\
     & -0.04370 & -0.0060 & 24 & 0.17379(45)  & 0.26928(37) & 0.8269(33)  & 1.1205(24) \\
     & -0.04630 & -0.0120 & 32 & 0.14399(59)  & 0.24358(59) & 0.8051(100) & 1.1210(46) \\
     & -0.04800 & -0.0023 & 32 & 0.13538(56)  & 0.26494(55) & 0.8297(59)  & 1.1631(62) \\
     & -0.04900 & -0.0060 & 32 & 0.12089(83)  & 0.25090(70) & 0.8104(77)  & 1.1679(84) \\
     & -0.04900 & -0.0120 & 32 & 0.11792(56)  & 0.23549(46) & 0.7937(64)  & 1.1458(76) \\
     & -0.05150 & -0.0120 & 48 & 0.08472(50)  & 0.22578(62) & 0.7613(78)  & 1.1846(107)\\
     & -0.05294 & -0.0060 & 64 & 0.06121(62)  & 0.23578(65) & 0.7877(49)  & 1.2211(147)\\\hline
3.61 & -0.02000 & -0.0042 & 32 & 0.19645(33)  & 0.23343(32) & 0.7020(60)  & 1.0503(08) \\
     & -0.02000 & -0.0045 & 32 & 0.19889(29)  & 0.25382(28) & 0.7204(51)  & 1.0724(10) \\
     & -0.02800 &  0.0045 & 32 & 0.14855(42)  & 0.23381(36) & 0.7051(48)  & 1.1249(26) \\
     & -0.03000 & -0.0042 & 32 & 0.12947(46)  & 0.20528(43) & 0.6642(53)  & 1.1126(32) \\
     & -0.03000 &  0.0045 & 32 & 0.13221(39)  & 0.22734(34) & 0.6926(43)  & 1.1441(35) \\
     & -0.03121 &  0.0045 & 32 & 0.12094(23)  & 0.22398(25) & 0.6986(40)  & 1.1456(28) \\
     & -0.03300 &  0.0045 & 48 & 0.10271(46)  & 0.21834(47) & 0.6786(49)  & 1.1629(61) \\
     & -0.03440 &  0.0045 & 48 & 0.08588(46)  & 0.21398(48) & 0.6828(49)  & 1.1997(74) \\
     & -0.03650 & -0.0030 & 64 & 0.04713(43)  & 0.18632(52) & 0.6383(54)  & 1.2280(103)\\\hline
3.70 & -0.00500 &  0.0500 & 32 & 0.22281(38)  & 0.32097(39) & 0.7952(54)  & 1.1268(30) \\
     & -0.01500 &  0.0000 & 32 & 0.16439(37)  & 0.20127(36) & 0.6128(116) & 1.0582(19) \\
     & -0.01500 &  0.0500 & 32 & 0.17104(35)  & 0.30039(35) & 0.7943(33)  & 1.1845(29) \\
     & -0.02080 & -0.0050 & 32 & 0.12498(38)  & 0.17172(34) & 0.5653(97)  & 1.0912(28) \\
     & -0.02080 &  0.0000 & 32 & 0.12464(51)  & 0.18368(47) & 0.5841(113) & 1.1007(36) \\
     & -0.02080 &  0.0010 & 32 & 0.12514(46)  & 0.18675(45) & 0.5678(68)  & 1.1177(38) \\
     & -0.02540 & -0.0050 & 48 & 0.08035(30)  & 0.15519(28) & 0.5392(55)  & 1.1348(54) \\
     & -0.02540 &  0.0000 & 48 & 0.08166(29)  & 0.16987(30) & 0.5585(43)  & 1.1606(54) \\
     & -0.02700 &  0.0000 & 64 & 0.06038(32)  & 0.16376(33) & 0.5569(57)  & 1.1649(68) \\\hline
3.80 & -0.00900 &  0.0000 & 32 & 0.15246(35)  & 0.17394(33) & 0.5116(94)  & 1.0453(14) \\
     & -0.01400 &  0.0000 & 32 & 0.12053(57)  & 0.15893(51) & 0.5096(122) & 1.0908(32) \\
     & -0.01400 &  0.0030 & 32 & 0.12304(54)  & 0.16818(52) & 0.4797(64)  & 1.0939(33) \\
     & -0.01900 &  0.0000 & 48 & 0.08200(86)  & 0.14461(90) & 0.4626(43)  & 1.1398(127)\\
     & -0.01900 &  0.0030 & 48 & 0.08230(105) & 0.15382(64) & 0.4743(65)  & 1.1819(145)\\
     & -0.02100 &  0.0000 & 64 & 0.05984(22)  & 0.13947(39) & 0.4658(33)  & 1.1657(113)\\\hline
\end{longtable}

The above table provides information about the individual ensembles for one choice of the extraction details.
The first four columns contain the input parameters for each simulation, i.e.\ the gauge coupling $\beta$, the values of the bare mass parameters $am_{\rm ud}^{\rm bare}$, $am_{\rm s}^{\rm bare}$ for the light and strange quarks, respectively, and the spatial extent in lattice units $L/a$ (the temporal extent of the lattices is typically larger and not listed here).
Note that the bare mass parameters can be negative, due to the additive quark mass renormalization with Wilson-type fermions, and the resulting (renormalized) quark masses are still positive.
The remaining columns show the quantities measured on these ensembles which are relevant for this work, and their statistical uncertainty as determined from the bootstrap procedure.
We like to stress that this table does not show the information on the correlation between measurements on the same ensembles, but of course by using the bootstrap samples for each quantity these correlations are properly taken into account in our analysis.
Further details on how these quantities were extracted can be found in Ref.~\cite{Durr:2013goa}.

\bigskip



\end{document}